\documentclass[aps,pre,twocolumn,showpacs,superscriptaddress]{revtex4}

\usepackage{epsfig}
\usepackage{amsmath}

\newcommand{\be}{\begin{equation}}
\newcommand{\ee}{\end{equation}}
\newcommand{\bc}{\begin{center}}
\newcommand{\ec}{\end{center}}
\newcommand{\bi}{\begin{itemize}}
\newcommand{\ei}{\end{itemize}}
\newcommand{\ba}{\begin{eqnarray}}
\newcommand{\ea}{\end{eqnarray}}

\begin{document}

\title{Structural and functional networks in complex systems with delay}

\author{V\'{\i}ctor M. Egu\'{\i}luz}
\email[Email: ]{victor@ifisc.uib-csic.es}
\affiliation{Instituto de F\'isica Interdisciplinar y Sistemas Complejos IFISC (CSIC-UIB), E07122 Palma de Mallorca, Spain}
\homepage{http://ifisc.uib-csic.es/victor}
\author{Toni P\'{e}rez}
\affiliation{Physics Department, Lehigh University, Bethlehem, PA 18015. USA}
\author{Javier Borge-Holthoefer}
\affiliation{Departament d'Enginyeria Inform\`atica i Matem\`atiques, Universitat Rovira i Virgili, 43007 Tarragona, Spain}
\author{Alex Arenas}
\affiliation{Departament d'Enginyeria Inform\`atica i Matem\`atiques, Universitat Rovira i Virgili, 43007 Tarragona, Spain}

\begin{abstract}

Functional networks of complex systems are obtained from the analysis of the temporal activity of their components, and are often used to infer their unknown underlying connectivity. We obtain the equations relating topology and function in a system of diffusively delay-coupled elements in complex networks. We solve exactly the resulting equations in motifs (directed structures of three nodes), and in directed networks. The mean-field solution for directed uncorrelated networks shows that the clusterization of the activity is dominated by the in-degree of the nodes, and that the locking frequency decreases with increasing average degree. We find that the exponent of a power law degree distribution of the structural topology, $\gamma$, is related to the exponent of the associated functional network as $\alpha =(2-\gamma)^{-1}$, for $\gamma < 2$.

\end{abstract}

\pacs{89.75.Fb, 05.40.-a, 05.65.+b, 89.75.Hc}
\maketitle
\section{Introduction}

Collective phenomena in populations of interacting elements is a subject of intense study in physical, biological, chemical, and social systems \cite{pikov,strogatz,physrep}. In many cases, the emergence of patches of coherent behavior is the main observable we have of the underlying dynamics and interaction of their constituents.
This is the case, for example, in gene expression, measured as DNA levels in microarrays \cite{stuart, mao}, or electrophysiological activity in the brain \cite{Siapas05}, measured through multi-unit extracellular electrode.
In many occasions, the coordination is not global but local, and the observation reveals clusters of elements dynamically correlated or, generally speaking, synchronized \cite{Gonzalez07,Danino10,Chavez10,Gregor10}. The resulting networks of coordinated activity are usually called functional networks of the system \cite{Bullmore09,Eguiluz05}. The analysis of these networks, from the physicists perspective, commonly focuses on network synchrony in the absence of time delays. However, delays are common in neural networks \cite{dha}, and many other biological \cite{tsai,Morelli09,Herrgen10} and social systems where the interaction between elements involves the propagation through a communication channel. The consideration of these delays is of utmost importance \cite{yu}. Recently, it has been analytically, and experimentally shown that zero time-lag synchronization is feasible over two distant (delayed) interacting oscillators when a third oscillator is placed in between of them \cite{fischer,vicente}. A recent study sheds light along these lines by studying the synchronization of networks of chaotic units with time-delayed couplings using the formalism of the master stability function \cite{kinzel}.

There has been a big effort from the scientific community to infer the complex network of interactions between elements from the functional network \cite{gerstein85,palm88,zhou06, yu06,timme07,pajevic08,napoletani08,barzel09}. Here we investigate on this inference from a fundamental physical perspective. We analyze the functional network resulting from the simplest dynamical system with delay presenting a synchronous dynamics, on a given topology, and relate topology and functionality. Given the simplicity of the model, we obtain the exact solution, develop a statistical mean-field theory approximation and find the relation between the degree distribution of the topological network and the associated functional network.

\section{The model}

First, we develop the analytical aspects of the problem. Let's start considering a set of $N$ elements coupled diffusively with delay
\be \dot{\phi_i} (t) = \omega_i +\epsilon \sum_j a_{ij} (\phi_j
(t-\tau_{ij}) - \phi_i(t))~, \label{set}
\ee
where $a_{ij}$ are the components of the adjacency matrix ${\bf A}$, that is, $a_{ij}=1$ if element $j$ influences $i$ with a delay $\tau_{ij}$, and $\epsilon$ is the coupling strength. Without loss of generality we can rescale the coupling strength to $\epsilon=1$. Equation~(\ref{set}) represents a system of $N$ elements in a network, which move at constant speed and that adjust their local position to match that of their neighbors; the communication between pair of nodes is not instantaneous but it is characterized by some delay. An alternative interpretation of Eq.~(\ref{set}) corresponds to the linearization of many non-linear interaction models, including the Kuramoto model \cite{Kuramoto1975}, as long as the phase differences remain small enough. As we will demonstrate later, if the network can be reached from at least one node, Eq.~(\ref{set}) presents a unique phase locked solution of the form
\be
\phi_i(t) = \Omega t + \theta_i~,
\label{phasesol}
\ee
where $\Omega$ is the locking frequency and $\theta_i$ the initial phase of element $i$. Substituting in Eq.~(\ref{set}), we obtain a set of $N$ linear equations that can be written in matrix form as
\be {\boldsymbol\omega} - \Omega \left( {\bf 1}+ {\bf T} \right) = {\bf L \boldsymbol\theta}~,
\label{grl}
\ee
where ${\bf L}$ is the Laplacian matrix  defined as
${L}_{ij}=k_{i,{\rm in}} \delta_{ij} - a_{ij}$,
$k_{i,{\rm in}} = \sum_j a_{ij}$ is the in-degree of node $i$,
$\delta_{ij}$ is the Kronecker delta,
{\bf 1} is a vector of 1's and
{\bf T} is a vector of components $T_i=\sum_j a_{ij}\tau_{ij}$ the total delay affecting each node,
$\boldsymbol\omega$ and $\boldsymbol\theta$ are the frequency vector and the phase vector with components $\omega_i$ and $\theta_i$ respectively.

\begin{figure*}
\includegraphics[width=0.9\linewidth]{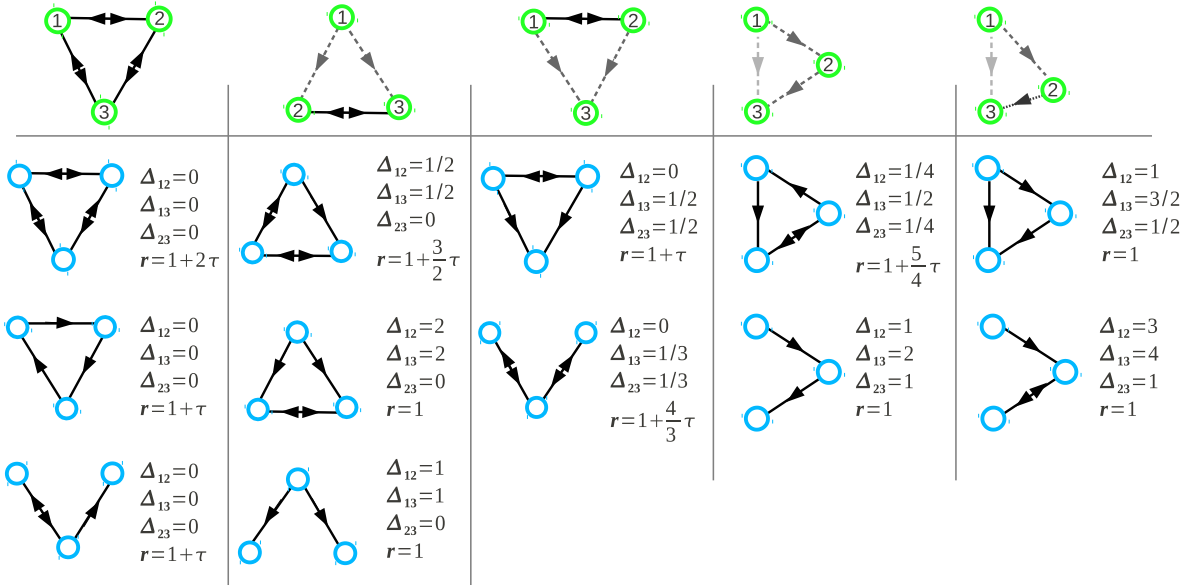}
\caption{\label{motifs} Functional network (green) for twelve structural motifs of three elements. In the functional solution the black solid
line represents a zero phase difference between the oscillators while the gray dashed lines stand for a non-zero phase difference. For the
structural, solid black lines stand for interaction links and directionality is indicated by an arrow. Parameters: $\Delta_{ij}=(\theta_i - \theta_j)/\Omega \tau$, and $r=\omega/\Omega$.}
\end{figure*}

In general, the Laplacian matrix ${\bf L}$ is asymmetric, however, as the sum of its rows is zero,
it admits a left-eigenvector $\mathbf{c}=\left(c_1, c_2, ..., c_N\right)$ with eigenvalue 0, that is, $\mathbf{c} {\bf L}=0$. The left eigenvector $\mathbf{c}$ is unique as long as all the nodes of the network can be reached by at least one node \cite{Chung:1997}. Left-multiplying Eq.~(\ref{grl}) by $\mathbf{c}$ we obtain the locking frequency
\begin{equation}
\Omega= \frac{\langle \boldsymbol\omega \rangle}{1+\langle\bf{T}\rangle}~,
\label{f1}
\end{equation}
where $\langle {\bf x}\rangle = \sum_i c_i x_i$ and $\mathbf{c}$ is normalized, $\sum_i c_i =1$. The phases are now given by
\be
\frac{\omega_i - \langle \boldsymbol\omega \rangle + \left( \omega_i \langle {\bf T} \rangle -
\langle \boldsymbol\omega \rangle T_i \right)}{1 + \langle {\bf T} \rangle} = ({\bf L \boldsymbol\theta})_i~.
\label{bid}
\ee
For undirected networks, the left-eigenvector $\mathbf{c}=\left(1/N, 1/N, ..., 1/N\right)$ is unique if it contains a single component and the brackets in Eq.~(\ref{f1}) are unweighted averages.

Thus, \emph{if all the nodes of the network can be reached by at least one node}, then Eq.~(\ref{set}) has a unique phase locked solution given by Eqs.~(\ref{f1}) and (\ref{bid}). These equations point out the relationship between the topology of the network, the distribution of delays, the locking frequency, and the state of the elements.

\subsection{Perfect synchronization}

The topology, delays and frequency can be combined to achieve the same state for each node. The condition for fully clustered solutions ($\theta_i = \theta_j$, $\forall i, j$) implies
\be
\frac{\omega_i}{\left( 1 + T_i \right)} = \Omega~.
\label{homozp}
\ee
From Eq.~(\ref{homozp}) we see that in absence
of delays ($T_i=0$) all the elements must have the same natural frequency, and reversely, if all nodes have the same natural frequency
($\omega_i=\omega$), the total delay affecting each node must be the same. It is straightforward to prove that Eq.~(\ref{homozp}) is
satisfied for degree regular networks of identical elements with equal delays, {\it i.e.}, $\omega_i=\omega$ and $\tau_{i}=\tau$. In this
case, the frequency of the phase synchronized state is $ \Omega = \frac{\omega}{1+\tau  k}$. In general, it is always possible to choose
the frequencies, the topology and the delays such that perfect coordinated activity is reached. In this case, the functional network, that
is, the network formed connecting those nodes displaying correlated activity, will be a fully connected network despite the sparse
connectivity of the underlying interaction network.

\subsection{Motifs}

Beyond the above perfectly synchronized solutions, we will extend our analysis to directed networks paying attention to the clusterization of the activity with different values of the phases. The simplest possible case corresponds to graphs of three nodes, motifs \cite{alon,motifbio,tsai}. The interest in motifs comes from previous studies showing the impact of motifs synchronization in absence of delays as building blocks of larger synchronized structures \cite{yamir}. Using Eq.~(\ref{grl}), we find the locked solution for each of the twelve different motifs of three elements with directed couplings. For the sake of clarity, we assume each element has the same frequency $\omega_i=\omega$,  and delay $\tau_{ij}=\tau$, $\forall i$, $j$. Solving Eq.~(\ref{grl}) we obtain for every motif configuration the normalized oscillation frequency $r^{-1}=\Omega/\omega$ and the phase differences $\left( \theta_i - \theta_j \right)=\Omega \tau \Delta_{ij}$.  The twelve different motifs are classified in five different functional networks (see Fig.~\ref{motifs}). This result points out the impossibility of deriving the motif topology solely from the information of the functional networks due to the degeneracy shown \cite{Kurths}.

\section{Heterogeneous mean field approach}

Beyond the formal exact solution presented in Eqs.~(\ref{f1})-(\ref{bid}), we want to gain
insight on the class of uncorrelated directed networks. First we start considering heterogeneous directed networks, specified by their
degree distribution $P({\bf k})$, where ${\bf k}=\left( k_{\rm in}, k_{\rm out} \right)$, and by the conditional probability
$P({\bf k}'|{\bf k})$ that a node of degree ${\bf k}$ is connected to a node of degree ${\bf k}'$. Normalization conditions
$\sum_{\bf k} P({\bf k}) = 1$ and $\sum_{{\bf k}'} P({\bf k}'|{\bf k}) = 1$ must be fulfilled. The degree detailed balance condition
$k_{\rm out}P({\bf k})P_{\rm out}({\bf k}'|{\bf k}) = k'_{\rm in}P({\bf k}')P_{\rm in}({\bf k}|{\bf k}')$ (where
$P_{\rm in}({\bf k}'|{\bf k})$ [$P_{\rm out}({\bf k}'|{\bf k})$] measures the probability to reach a vertex of degree ${\bf k}'$ leaving from
a vertex of degree ${\bf k}$ using an incoming [outgoing] edge of the source vertex) ensures that the network is closed and that
$\langle k_{\rm in}\rangle = \langle k_{\rm out} \rangle$.
We resort on the heterogeneous mean-field approach, coarse-graining the dynamics to classes of nodes of the same degree ${\bf k}$.
Thus, we define the phase density $\Phi_{\bf k}$ of nodes of degree ${\bf k}$ as
\be
\Phi_{\bf k} = \frac{1}{N_{\bf k}} \sum_{i\in K} \phi_i~,
\ee
where $N_{\bf k}=P({\bf k}) N$ is the expected number of nodes with degree ${\bf k}$. Here we have made use of $K$ to denote the set of nodes with degree ${\bf k}$. Similarly we define the frequency density
\be
W_{\bf k} = \frac{1}{N_{\bf k}} \sum_{i\in K} \omega_i~.
\ee
This notation allows to group the sums by the degrees of the nodes. For instance, if the degree of node $i$ is ${\bf k}_i={\bf k}$ then
\be
\sum_j a_{ij} \phi_j = k_{\rm in} \sum_{{\bf k}'} P_{\rm in}({\bf k}'|{\bf k})
\Phi_{{\bf k}'}~.
\ee

For identical elements $\omega_i = \omega$ and $\tau_{ij}=\tau$, the time
evolution of the phase density of the class of nodes of degree ${\bf k}$, $\Phi_{\bf k}(t)$, can be rewritten from Eq.~(\ref{set}) as
\be
\dot{\Phi}_{\bf k}(t) = W_{\bf k} +k_{\rm in}\sum_{{\bf k}'} P_{\rm in}({\bf
k}'|{\bf k}) (\Phi_{{\bf k}'}(t-\tau) - \Phi_{\bf k}(t))~. \label{hmf}
\ee

For uncorrelated networks $P_{\rm in}({\bf k}'|{\bf k}) =\frac{k_{\rm out}' P({\bf k}')}{\langle k\rangle}$ and with the ansatz of locked
solutions $\Phi_{\bf k} = \Omega t + \Theta_{\bf k}$, we obtain
\be
\Omega = W_{\bf k} - k_{\rm in} \Omega \tau +  \frac{k_{\rm in}}{\langle k
\rangle} \sum_{{\bf k}'}
k_{\rm out}' P({\bf k}') (\Theta_{{\bf k}'} - \Theta_{\bf k})~.
\ee
Summing over all degrees we find
\ba
\Theta_{\bf k} &=& \frac{\Omega \langle k \rangle \tau}{\langle \frac{k_{\rm
out}}{k_{\rm in}}\rangle}
\frac{1}{k_{\rm in}} + a ~,\\
\Omega &=& \frac{\omega}{1 + \frac{\langle k \rangle}{\langle
k_{\rm out}/k_{\rm in}\rangle} \tau}~,
\ea
being $a$ an arbitrary constant. For undirected networks $k_{\rm in}=k_{\rm out}$, thus we recover Eq.~(\ref{f1}) for the locking frequency
%
where $\langle k \rangle \tau = \langle {\bf T} \rangle$, and
\be
\Theta_k = \frac{\Omega \langle k \rangle \tau}{k} + a~.
\label{Thetak}
\ee
This indicates that whether two nodes show a similar phase depends on their degree difference in an uncorrelated network. It also shows that low-degree nodes are ahead of high-degree nodes. At least in this limit, the precise shape of the degree distribution is not playing an important role, as only the average degree $\langle k \rangle$ enters into the equation. Obviously, this dependence of the degree is reminiscent of our hypothesis of a mean-field coarse-grained by degree, however it is not trivial that this approximation will hold for the actual dynamics (Eq.~(\ref{set})).

\begin{figure}
\includegraphics[width=0.9\linewidth]{Fig2a.eps}
\includegraphics[width=0.9\linewidth]{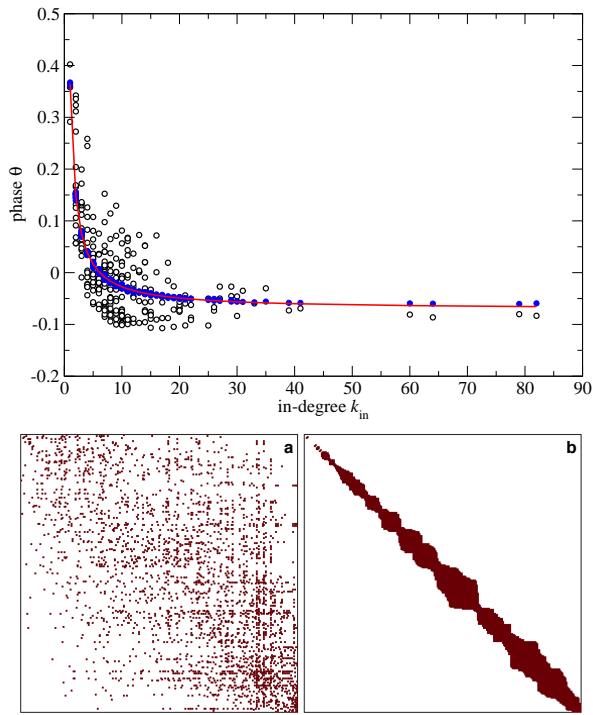}
\caption{\label{f:ce_sina} (Top row) Phase clusterization in the directed {\em C.\ elegans} neural network. We plot the phases versus the in-degree $k_{\rm in}$ for neurons in the {\em C.\ elegans} (open circles) and rewired networks (filled symbols) (averaged over 100 realizations of the rewiring algorithm) keeping the same $(k_{\rm in}, k_{\rm out})$ for each neuron. In the rewired networks the phase is well approximated by the relation $\Theta(k)= b/k + a$. The phases are obtained after integration of Eq.~(1) with $\omega_i = 1$ and $\tau_{ij} =0.1$.
(Bottom row) Adjacency matrix of the neuronal connections (a), and of the functional network (b). In both cases the neurons are ordered according to the ranking of their phases obtained from the dynamical system given by Eq.~(1) with $\omega_i =1$ and $\tau_{ij} = 0.1$.
}
\end{figure}

Further more the distribution of the phases in correlated networks as for example in the {\em C.\ elegans} neural network also shows a good agreement with Eq.~(\ref{Thetak}). The neuronal network connectivity of the {\em C.\ elegans} can be represented as a weighted adjacency matrix of 275 nonpharyngeal neurons, out of a total of 302 neurons (http://www.wormatlas.org/). We assume that the nervous system of the {\em C.\ elegans} can be modeled as a network, where nodes represent the center of the cell bodies, and the links represent synapses. The heterogeneous mean field formalism describes the relationship between dynamics and topology in uncorrelated networks. Such relationship can be illustrated in a real (correlated) network analyzing the dynamics of Eq.~(1) using the connectivity of the neural system of the {\em C.\ elegans} (Fig.~\ref{f:ce_sina}). When comparing the exact solution in the directed neural network with the analytical solution we observe that it captures the dependence on the in-degree and gives an excellent solution for the rewired directed network. Thus the in-degree of a neuron gives a good first approximation to the real state of the neuron although the precise wiring details are very important to know its exact value. Simple models aiming at the reconstruction of the anatomical network based on the observed neurons' states will link, for this dynamics, neurons with similar in-degree with no connection in the real network (see Fig.~\ref{f:ce_sina}).

\begin{figure}[t]
\includegraphics[width=\linewidth]{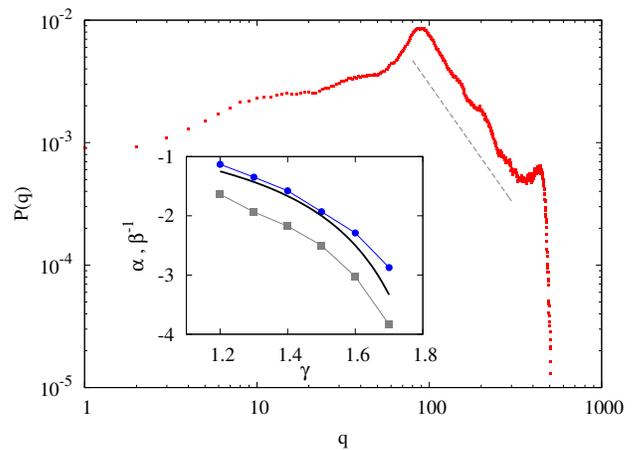}
\caption{\label{f:expos} Functional degree distribution $P(q)$ for a network of $10^5$ nodes with structural degree distribution
$P(k) \sim k^{-\gamma}$ and $\gamma=1.5$. Dashed line corresponds to $q^{-2}$. Inset: Dependence of $\beta^{-1}$ (circles) and $\alpha$
(squares) on the structural degree distribution exponent $\gamma$. Solid black line represents the theoretical prediction. The functional
networks are obtained after integration of Eq.~(\ref{set}) with $\omega_i = 1$ and $\tau_{ij} =0.1$, and $\Delta=10^{-4}$.}

\end{figure}

The heterogeneous mean field solution allows us to relate the degree distributions of the structural and functional topologies. In the remainder we will assume undirected structural networks. In the functional network a node with degree $k$
is connected with a node with degree $k'$ if their phase difference is smaller than a given threshold: $|\Phi_k - \Phi_{k'}|\le \Delta$.
Then, using Eq.~(\ref{Thetak}), the functional degree of a node of structural degree $k$ is given by $q(k) \sim \int_{|1/k - 1/k'| \le \delta} P(k') dk'$
where $\delta$ is an arbitrary threshold $\delta \omega \langle k \rangle \tau = \Delta (1 + \langle k \rangle \tau)$). If the degree distribution of the structural network is a power law
$P(k)\sim k^{-\gamma}$ then, $q(k)\sim k^\beta$ where $\beta = 2-\gamma$ for $\gamma < 2$ and the degree distribution of the functional
network is also power law $P(q)\sim q^{-\alpha}$ where $\alpha = (2-\gamma)^{-1}$. The numerical simulations of the system given by
Eq. (\ref{set}) shows an excellent agreement with the analytical prediction for classes of nodes with degree $k$ in uncorrelated networks.
In Fig.~\ref{f:expos} we compare the values of the exponent $\alpha$ and $\beta$ obtained after integration of the dynamical system given by Eq.~(\ref{set}) in scale-free
networks.

\section{Conclusions}

Summarizing, we have got insight in the relationship between the topological network of connections and the functional network obtained from a simple dynamical process with delays. We have found the conditions for the emergence of locked dynamical states in any network of diffusively delay-coupled oscillators. We identify these states as the main components of the emergent functional network generated by this simplified dynamics. Using these analytical guides we have explored the functional network obtained for the class of uncorrelated heterogenous networks, under the mean-field hypothesis, and have checked its prediction in scale-free networks. The results allow us to grasp the dependence of the functional network on the topological parameters, highlighting the role played by the delays and heterogeneity \cite{brain}. Indeed, although functional and structural topologies differ at the local level, we have shown that the degree distributions are related in the presence of delays as distant nodes sharing the same degree will be functionally correlated.

\section{Acknowledgments}
V.M.E., A.A. and J.B.-H. acknowledge financial support from MEC (Spain) through projects FISICOS (FIS2007-60327) and FIS2009-13730-C02-02. T.P. acknowledges support from the NSF, and Mathers Foundation. A.A. acknowledges partial support by the Director, Office of Science, Computational and Technology Research, U.S. Department of Energy under Contract No. DE-AC02-05CH11231.



\end{document}